\documentclass[aps,prl,twocolumn,superscriptaddress]{revtex4-1}
\usepackage{graphicx}
\usepackage{bm}
\usepackage{subfigure}

\usepackage{xspace}
\usepackage{mathrsfs}
\usepackage[usenames,dvipsnames]{color}

\newcommand{\HH}{\ensuremath{\mathscr{H}}}
\newcommand{\smmu}{\ensuremath{\mu_{\mathrm{s}}}\xspace}
\newcommand{\smJij}{\ensuremath{J_{ij}}\xspace}

\newcommand{\smKu}{\ensuremath{k_{\mathrm{u}}}\xspace}

\newcommand{\sms}{\ensuremath{\mathbf{S}}\xspace}

\newcommand{\vampire}{\textsc{vampire}\xspace}

\newcommand{\muB}{\ensuremath{\mu_{\mathrm{B}}}\xspace}

\begin{document}

\title{Ultrafast thermally induced magnetic switching in synthetic ferrimagnets}

\author{Richard F. L. Evans}
\email{richard.evans@york.ac.uk}
\author{Thomas A. Ostler}
\author{Roy W. Chantrell}
\affiliation{Department of Physics, University of York, Heslington, York YO10 5DD United Kingdom.}
\author{Ilie Radu}
\affiliation{Institut f\"ur Methoden und Instrumentierung der Forschung mit Synchrotronstrahlung, Helmholtz-Zentrum Berlin f\"ur Materialien und Energie, GmbH, Albert-Einstein-Stra\ss e 15,
12489 Berlin, Germany}
\author{Theo Rasing}
\affiliation{Radboud University Nijmegen, Institute for Molecules and Materials, Heyendaalseweg 135, 6525 AJ Nijmegen, The Netherlands.}

\begin{abstract}
Synthetic ferrimagnets are composite magnetic structures formed from two or more anti-ferromagnetically coupled magnetic sublattices with different magnetic moments. Here we report on atomistic spin simulations of the laser-induced magnetization dynamics on such synthetic ferrimagnets, and demonstrate that the application of ultrashort laser pulses leads to sub-picoscond magnetization dynamics and all-optical switching in a similar manner as in ferrimagnetic alloys. Moreover, we present the essential material properties for successful laser-induced switching, demonstrating the feasibility of using a synthetic ferrimagnet as a high density magnetic storage element without the need of a write field.
\end{abstract}

\maketitle
The dynamic response of magnetic materials to ultra-short laser pulses is currently an area of fundamental and practical importance that is attracting a lot of attention. Since the pioneering work of Beaurepaire\textit{et al} \cite{BeaurepairePRL1996} it has been known that the magnetization can respond to a femtosecond laser pulse on a sub-picosecond timescale. However studies of magnetic switching are more recent. In this context an especially intriguing phenomenon is that of all-optical switching, which uses the interaction of short, intense pulses of light with a magnetic material to alter its magnetic state without the application of an external magnetic field\cite{KimelAOS2005,StanciuAOS2007}. Recent experiments \cite{RaduNature2011,OstlerNatCom2012,KhorsandMCD2012} and
theoretical calculations\cite{OstlerNatCom2012,AtxitiaUFRev2013,MentinkPRL2012,BarkerSR2013} have demonstrated that the origin of all-optical switching in ferrimagnetic alloys is due to ultrafast heating of the spin system. The magnetic switching arises due to a transfer of angular momentum between the two sublattices within the material\cite{AtxitiaUFRev2013,MentinkPRL2012} and the resulting exchange-field induced precession\cite{AtxitiaUFRev2013}. Remarkably, this effect occurs in the absence of any symmetry breaking magnetic field \cite{OstlerNatCom2012}, and can be considered as Thermally Induced Magnetic Switching (TIMS). So far TIMS has only been demonstrated experimentally in the rare-earth transition metal (RE-TM) alloys GdFeCo and TbCo which, in addition to their strong magneto-optical response, have two essential properties for heat-induced switching: antiferromagnetic coupling between the RE and TM sublattices\cite{OstlerGdFeCoPRB2011} and distinct demagnetization times of the two sublattices\cite{RaduNature2011}. The antiferromagnetic coupling allows for inertial magnetization dynamics, while the distinct demagnetization times under the action of a heat pulse allow a transient imbalance in the angular momentum of the two sublattices, which initiates a mutual high speed precession enabling ultrafast switching to occur.

\begin{figure}[!t]
\includegraphics[width=7cm]{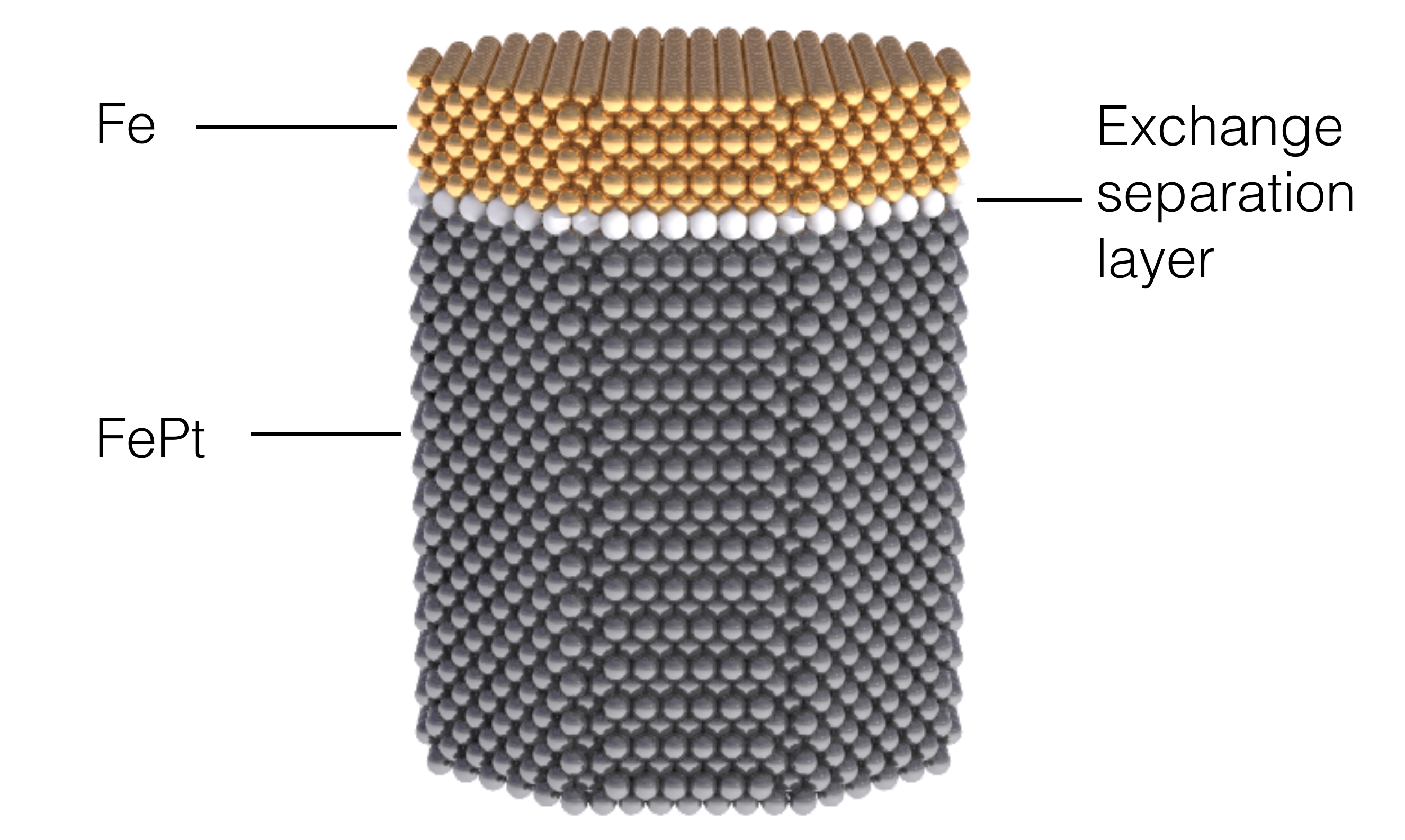}
\caption{Visualization of a synthetic ferrimagnetic structure consisting of two ferromagnetic layers separated by a non-magnetic spacer layer to engineer anti-ferromagnetic coupling between them. (Color Online).}
\label{fig:schematic}
\end{figure}

Although GdFeCo has excellent switching properties, its potential use in magnetic data storage is limited by its low anisotropy and amorphous structure, precluding the use of single magnetic domains typically less than 10 nm in size, required for future high density magnetic recording media. One intriguing possibility, and the focus of this paper, would be the use of a synthetic ferrimagnet (SFiM), consisting of two transition metal ferromagnets anti-ferromagnetically exchange coupled by a non-magnetic spacer\cite{ParkinPRL1991}, shown schematically in Fig~\ref{fig:schematic}. The important but as yet unanswered question is whether all-optical switching would also work in such an artificial structure and what essential physical properties of the design are required. Such a composite magnet also has a number of distinct advantages over intrinsic rare-earth-transition metal ferrimagnets: the dynamic properties of each sublattice may be separately selected by choice of material, nano-patterning is possible in the sub-10 nm size range due to their crystalline nature and the omission of costly rare-earth metals. Importantly the composite design has the advantage of allowing the use of high anisotropy materials such as FePt or CoPt to enhance the thermal stability of the medium. These advantages could make such synthetic structures very promising candidates for magnetic data storage applications.

In this letter we present dynamic studies of such a synthetic ferrimagnet using an atomistic spin model. We investigate the dynamic properties of the separate layers and show that the demagnetization time is determined primarily by the local atomic spin moment and the intrinsic Gilbert damping of the material. We finally consider an exchange-coupled Fe/FePt synthetic ferrimagnet and show that a short heat-pulse is sufficient to induce ultrafast heat-induced switching of the material.

The dynamic properties of the SFiM are studied using an atomistic spin model using the \vampire software package\cite{vampire-url,EvansVMPR2013}. The energetics of the system are described using a Heisenberg spin Hamiltonian, which in condensed  form reads:
\begin{equation}
\HH = -\sum_{i<j} \smJij \sms_i \cdot \sms_j -\sum_{i} \smKu S_{i,z}^2
\label{Hamiltonian}
\end{equation}
where \smJij is the exchange energy between nearest neighboring spins, $\sms_i$ and $\sms_j$ are unit vectors describing the spin directions for local sites $i$ and nearest neighbor sites $j$ respectively, and \smKu is the uniaxial anisotropy constant. There are three distinct \smJij exchange interactions in the SFiM system, arising from the intralayer and interlayer contributions, detailed in Tab.~\ref{tab:parameters}. Since we are considering heat-induced switching no external field is applied during the simulations. The system of coupled spins is integrated using the Landau-Lifshitz-Gilbert equation with the Langevin dynamics formalism at the atomistic level using the Heun numerical scheme\cite{EvansVMPR2013}. More details of the model are provided in the Supplementary Information\cite{SuppInfo}.

For a synthetic ferrimagnet to exhibit thermally induced magnetic switching it is essential to consider the physical requirements of the structure analogous to those of intrinsic RE-TM ferrimagnets. The first property is the anti-ferromagnetic exchange coupling of the component layers of the synthetic ferrimagnet, which can be engineered by a suitable choice and thickness of material such as Ir or Ru\cite{ParkinPRL1991}. The second criterion is the existence of distinct magnetization dynamics for the two component layers which allows the formation of a transient ferromagnetic state and drives the switching process\cite{RaduNature2011,OstlerNatCom2012}.

\begin{figure}[!t]
\includegraphics[width=7.8cm]{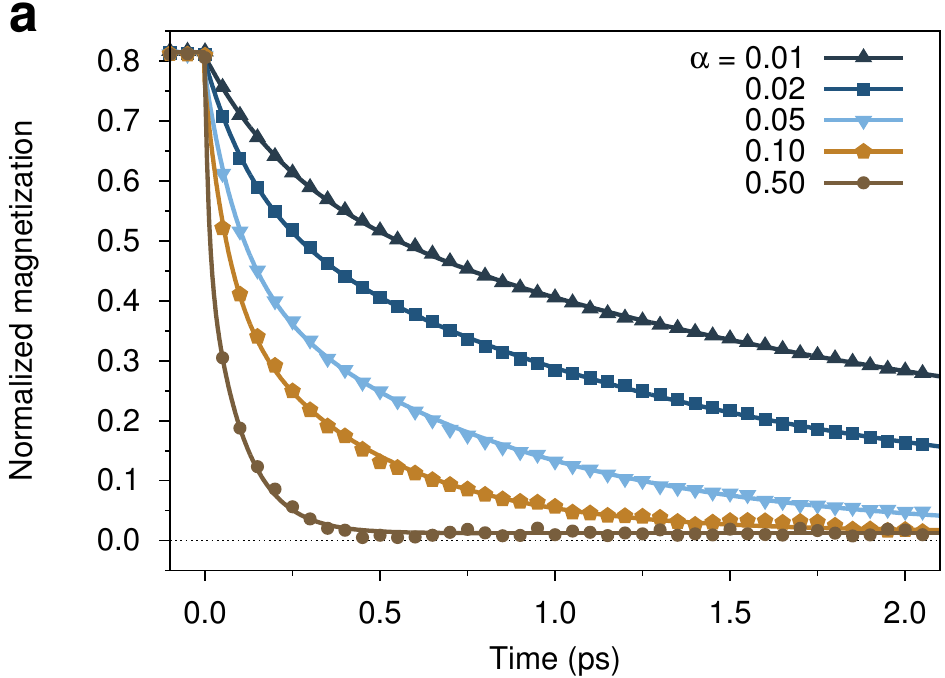}
\includegraphics[width=7.8cm]{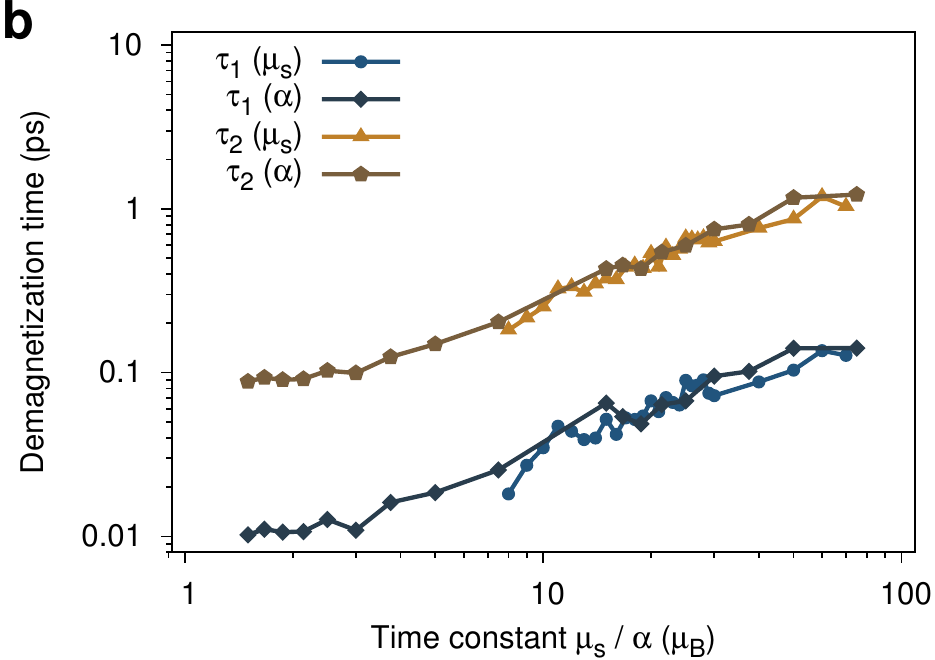}
\caption{(\textbf{a}) Simulated demagnetization of a ferromagnet under the action of a heat pulse (points) for different values of the intrinsic damping constant, $\alpha$, fitted to Eq.~\ref{eq:dmt} (lines). (\textbf{b}) Fitted demagnetization time constants $\tau_1$ and $\tau_2$ as a function of the ratio $\smmu/\alpha$ for a ferromagnetic material. The legend shows the origin of the calculated demagnetization time for variable $\alpha$ ($\smmu = 1.5 \muB$) or for variable \smmu ($\alpha=0.1$) respectively. The data show that the the demagnetization time constants for the same exchange interaction depend only on the ratio of $\smmu/\alpha$, and not the individual values of \smmu and $\alpha$. (Color Online).}
\label{fig:mu-alpha}
\end{figure}

Let us start with a simplistic scenario, where the effects of exchange are ignored and the demagnetization time $\tau_\mathrm{d}$ of a ferromagnetic material is given by the ratio of the atomic magnetic moment and damping\cite{KoopmansPRL2005}, such that $\tau_\mathrm{d}\sim \smmu/\alpha$. In this simplistic view, a variation of the damping parameter or the local atomic spin moment will lead to a straightforward variation of the demagnetization time. To test this assertion, we have simulated a \textit{ferromagnetic} material, with an exchange coupling the same as FePt, but with freely varied local atomic spin moment \smmu and damping parameter, $\alpha$, to which a step-function increase to a temperature above the Curie temperature is applied. The sudden increase in temperature leads to a demagnetization of the material, with characteristic behavior shown in Fig.~\ref{fig:mu-alpha}(a) for different values of the damping parameter. The demagnetization dynamics are not generally describable by a single demagnetization time $\tau_\mathrm{d}$, but usually it is sufficient to describe the demagnetization dynamics by two leading contributions\cite{Garanin}. We therefore fit the time-dependent demagnetization dynamics with the function\cite{KoopmansNatMat2010,RaduNature2011}
\begin{equation}
m(t) = A_1 \exp\left(-\frac{t}{\tau_1}\right) + A_2\exp\left(-\frac{t}{\tau_2}\right) + \mathrm{const.}
\label{eq:dmt}
\end{equation}
where the demagnetization time constants $\tau_1$ and $\tau_2$ are intrinsic timescales and $A_1$, $A_2$ and the constant are fitting parameters. Since the final temperature is above the Curie point the constant is close to zero but is treated as a free parameter of the fitting.

A systematic variation of the Gilbert damping parameter and local spin moment leads to a range of demagnetization times (by fitting to Eq.~\ref{eq:dmt}), shown in Fig.~\ref{fig:mu-alpha}(b). The essential result is that the demagnetization time constants, for a fixed exchange constant, depends only on the ratio $\smmu/\alpha$. Thus a certain demagnetization time can be engineered through any combination of the local spin moment (changeable by varying the composition in FeCo alloys for example), and the damping parameter (for example by using high anisotropy materials or by introducing disorder in the form of defects and impurities).

In addition to the requirements of antiferromagnetic coupling and distinct demagnetization times, the thickness of the two layers is also important for synthetic ferrimagnets. In RE-TM alloys the antiferromagnetic exchange interactions are a bulk effect existing between neighboring atoms. For the synthetic structure however the exchange is an interface effect, so the effective exchange field on the layer is inversely proportional to the layer thickness. Hence for large effective coupling thin layers are preferred. For thermal stability larger volumes are preferred, and so there must be a balance between this and the effective exchange field. In general it appears to be preferable for the high anisotropy material to have a larger volume, while the other layer can be thin to maximize the effective exchange field.

\begin{figure}[!t]
\includegraphics[width=7.8cm]{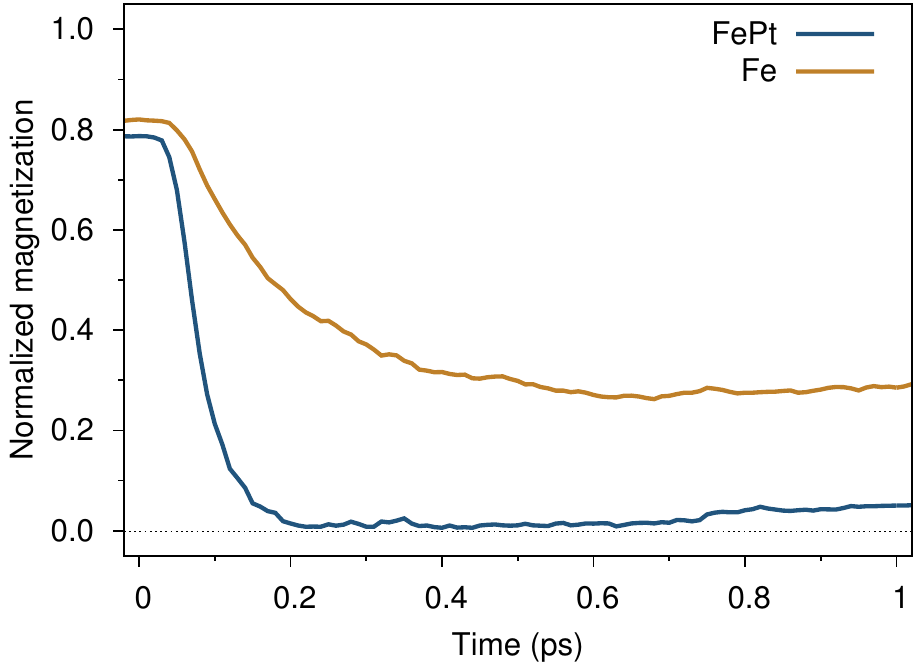}
\caption{Simulated magnetization dynamics for Fe and FePt ferromagnets under action of an ultrafast heat pulse Gaussian in time with a width of 20 fs. The magnetization is normalized to the 0K saturation value, and so the different Curie temperatures lead to different reduced magnetization at the starting temperature of T=300K. The heat pulse is sufficient to demagnetize the FePt layer, but the Fe layer remains ordered. (Color Online).}
\label{fig:demag}
\end{figure}

Having defined the essential physical parameters for thermally induced switching in synthetic ferrimagnets, we now consider specific materials which may be suitable for applications. For devices, strong magnetocrystalline anisotropy is essential, and so the obvious choice (excluding rare-earth metals) are CoPt and FePt based alloys, currently used in conventional magnetic recording media. The dynamic properties of such materials can be inferred from theoretical calculations of atomic moments and experimental results of the effective damping. In L$1_0$ FePt the magnetic moments are not evenly distributed between the Fe and Pt atoms, since pure Pt is non-magnetic\cite{MyrasovFePtEPL2005}. \textit{Ab-initio} calculations\cite{MyrasovFePtEPL2005} show that the Fe moments in FePt are significantly enhanced over bulk Fe, with an effective moment of $3.2$ \muB, and so one would normally assume relatively slow dynamics for the Fe moments in FePt. However, the strong spin-orbit coupling in FePt also leads to large Gilbert damping around 0.1\cite{ZhangFePtdampingAPL2010}, and so FePt as a whole exhibits `fast' dynamics. The relatively low Curie temperature of $\sim 700$K is also suitable for laser induced switching as the threshold fluence is lower. For heat-induced switching the two layers must exhibit distinct magnetization dynamics, and so the other layer must be `slower' than FePt. The obvious choices are elemental Fe or Co, due to their lower damping and high moments. Here we choose Fe due to its lower Curie temperature of 1043K, and larger magnetic moment of 2.2 \muB.

Before considering the switching properties of the Fe/FePt synthetic ferrimagnet, we first address the demagnetization dynamics of the two uncoupled layers individually. The system consists of a 5 nm thick FePt layer and 1 nm thick Fe layer with a lateral size of 8 nm in the shape of a cylinder. The temporal temperature profile arising from a 20 fs laser pulse is calculated using the two temperature model treating the electron and phonon systems separately and coupling the magnetic system to the electrons\cite{AnisimovTTM1974}. The parameters for the two temperature model are the same as those in Ref.~\onlinecite{OstlerGdFeCoPRB2011} and further details are provided in the Supplementary Information\cite{SuppInfo}. The resulting demagnetization dynamics for the Fe and FePt layers are presented in Fig.~\ref{fig:demag}. As expected, the FePt and Fe layers exhibit `fast' and `slow' dynamics respectively, but the different Curie temperatures of the two layers leads to different levels of demagnetization. 

\begin{figure}[!tb]
\includegraphics[width=7.8cm]{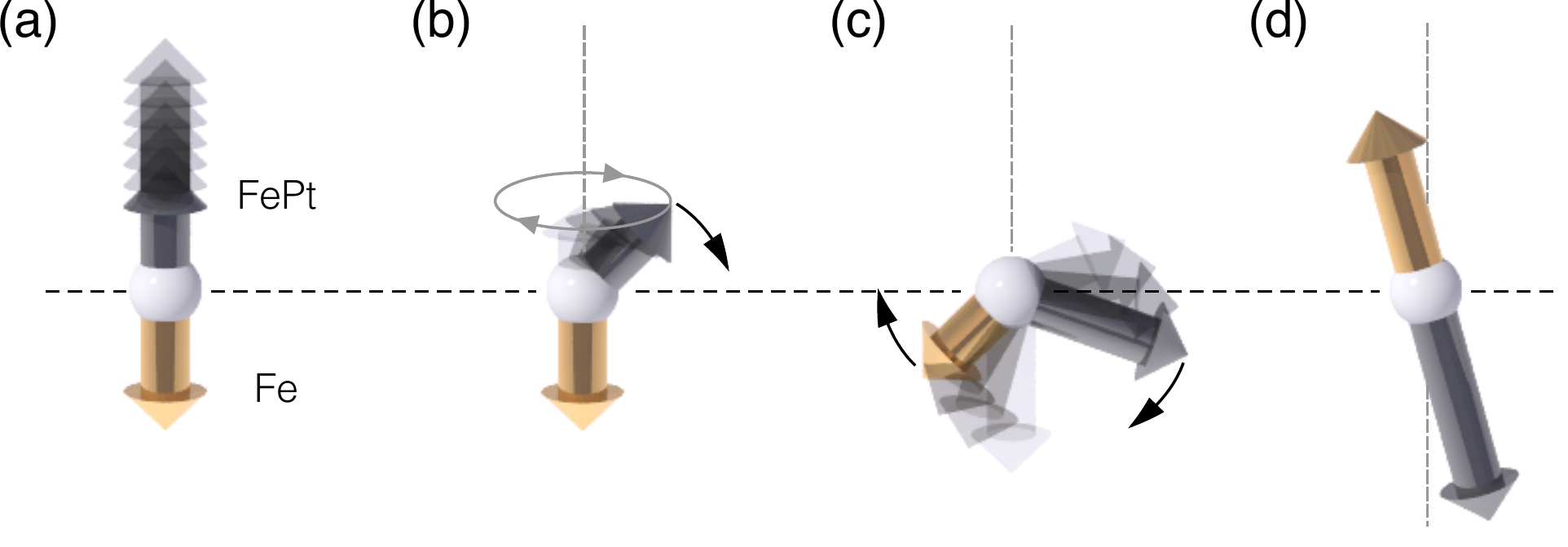}
\caption{Schematic illustration of the switching process for all-optical heat-induced magnetic switching. (\textbf{a}) The two sublattices are initially aligned anti-parallel, and application of a laser pulse causes rapid demagnetization of the FePt sublattice. (\textbf{b}) Thermal fluctuations lead to a small transverse component which initiates a rapid precession of the FePt sublattice in the exchange field of the Fe layer and relaxation toward the Fe sublattice. (\textbf{c}) As the FePt layer precesses the Fe layer responds to the laser pulse and begins mutual precession leading to a transient ferromagnetic state. (\textbf{d}) After the system has reached thermal equilibrium the sublattices are aligned antiparallel and relax to the easy axis direction, completing the switching process. (Color Online.)}
\label{fig:switching-schematic}
\end{figure}

\begin{figure}[!t]
\includegraphics[width=7.8cm]{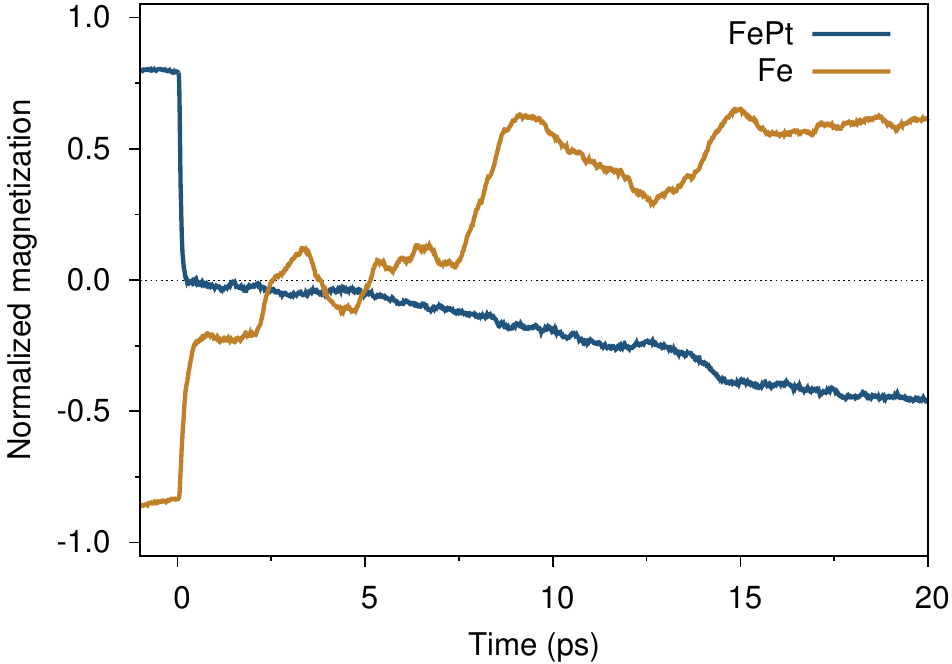}
\caption{Calculated magnetization dynamics for the Fe/FePt bilayer in response to an ultrafast heat pulse. The FePt layer rapidly demagnetizes and develops a transverse component, inducing rapid precession leading to a transient ferromagnetic state. The mutual precession of the Fe and FePt sublattices leads to switching of the Fe sublattice, while the FePt sublattice is stabilized by the high magnetocrystalline anisotropy. (Color Online).}
\label{fig:switching}
\end{figure}

We finally consider ultrafast all-optical switching in the SFiM nanostructure. Heat-induced switching is driven by an intricate process involving ultrafast demagnetization, transfer of angular momentum between the sublattices, and high frequency precession via the exchange mode\cite{AtxitiaUFRev2013}, shown schematically in Fig.~\ref{fig:switching-schematic}. The strength of the exchange interaction determines the timescale of the switching process and the rate of transfer of angular momentum between the sublattices and so should be as strong as possible. Depending on the system there is some disparity between the values of exchange measured experimentally\cite{ParkinPRL1991,JiangJAP2005} and calculated theoretically\cite{HerperPRB2002}, and so we assume an intermediate value of the interlayer exchange interaction of $\sim 1/5$ that of the bulk FePt exchange ($J^{\mathrm{FePt-Fe}}_{ij} = -1.635 \times 10^{-21}$ J/link). The calculated sublattice magnetization dynamics resulting from a 20 fs heat pulse is shown in Fig.~\ref{fig:switching}.

The FePt layer is initially demagnetized by the heat pulse and switches direction under the influence of the exchange field from the Fe, while the Fe layer itself retains a higher degree of order. As the electron system cools the FePt recovers its order along the -$z$ direction i.e. it switches its magnetization. Simultaneously the Fe sublattice begins to precess in the increasing exchange field of the FePt layer, and switches precessionally, as seen from the oscillations in the Fe magnetization in Fig.~\ref{fig:switching}. Finally the system relaxes to the usual (but reversed) antiferromagnetic state, with some final relaxation towards the easy axis direction on a longer timescale. The simulations demonstrate the feasibility of ultrafast heat-induced switching in synthetic ferrimagnetic nanostructures with the correct physical characteristics.

Although the switching is a thermally driven process, the mechanism is itself deterministic and quite robust. To demonstrate this, we present similar switching for a CoPt/Fe bilayer structure in the Supplementary Information\cite{SuppInfo}. The exchange coupling between the two layers is an essential component for the switching. For fast dynamics this should ideally be as large as possible, but switching also occurs for typical values of $\sim 6$ mJ/m$^2$ reported experimentally\cite{ParkinPRL1991}, as presented in the Supplementary Information\cite{SuppInfo}.

To conclude, we have investigated the laser induced dynamic properties of synthetic ferrimagnets, and demonstrated the possibility of all-optical magnetic switching in Fe/FePt nano structures. Material combinations other than Fe/FePt are possible, including other alloys based on Ni, Fe and Co ferromagnets such as Fe/CoPt. The interlayer exchange coupling, which drives the switching process, can also be engineered by using different materials such as Ir or Si, though the optimal coupling depends on the thickness of the layers. This allows the tuning of properties such as the Curie temperature, magnetocrystalline anisotropy and Gilbert damping to achieve the desired dynamic behavior and switching properties, and opens the possibility of engineering high performance magnetic data storage devices. This is highly significant for the following reasons. Firstly, the removal of the necessity for the write field in magnetic recording write transducers would lead to a dramatic reduction in the complexity of transducer design and the number of operations required for their production. Secondly, it has been shown that that the magnitude of the write field is a major factor limiting the ultimate density in magnetic recording\cite{EvansAPL2012}. Essentially, the field during writing must be sufficiently large to overcome thermally driven back-switching of the magnetization, a factor termed the thermal writability. In Ref.~\cite{EvansAPL2012} it was shown that thermal writability is a more important factor than the thermal stability criterion in determining the limiting recording density. The effective field in the TIMS process is the exchange field between the sublattices, which is significantly larger than the values accessible by today's inductive technology, which would essentially remove the thermal writability as a limiting factor in magnetic recording.

\begin{table}[!htb]
\caption{Summary table of model parameters and their units}
\begin{ruledtabular}
\begin{tabular}{ l c c l}
  &   Fe     & FePt & Unit            \\
  \cline{1-4}
  \smJij    & $6.75 \times 10^{-21}$  & 4.5 $\times 10^{-21}$  & J/link\\ 
  \smmu     &      2.2               &         3.2            & \muB\\
  \smKu     & 0.0 & $1.61\times 10^{-22}$  & J/atom \\
  $\alpha$ & 0.01                   & 0.1                    & - \\
\end{tabular}
\end{ruledtabular}
\label{tab:parameters}
\end{table}

Financial support from the EU Seventh Framework Programme under grant agreement No. 281043 \textsc{femtospin} is gratefully acknowledged.


%

\end{document}